\documentclass[a4paper,11pt]{article}
\pdfoutput=1

\usepackage{jcappub} 
\usepackage{natbib}
\setcitestyle{square,comma,numbers,sort&compress}
\usepackage{enumerate}
\usepackage{tabularx}
\usepackage[left = 1.0in, top=1.0in,right=1.0in,bottom=1.0in,a4paper]{geometry}
\usepackage[dvipsnames]{xcolor}
\usepackage{comment}
\usepackage[caption=false]{subfig}
\usepackage{cancel}

\usepackage[section]{placeins}
\usepackage{listings}
\usepackage{amsbsy}
\newcolumntype{Y}{>{\centering\arraybackslash}X}

\definecolor{myRED}{rgb}{0.8, 0.25, 0.33}
\usepackage[normalem]{ulem}
\usepackage{dsfont}
\usepackage{pbox}
\usepackage{graphicx}
\usepackage{multirow}
\usepackage{tikz}
\usetikzlibrary{arrows}
\usepackage{enumitem} 
\usepackage{ulem}
\usepackage{soul}
\usepackage{lipsum} 
\allowdisplaybreaks

\title{\boldmath\huge  
Ultraviolet Completion of a Two-loop Neutrino Mass Model}

\author[a]{K.S. Babu,}
\author[b]{Shaikh Saad}

\emailAdd{babu@okstate.edu, shaikh.saad@okstate.edu}

\affiliation[a]{Department of Physics, Oklahom State University, Stillwater, OK 74078, USA}

\affiliation[b]{Jožef Stefan Institute, Jamova 39, P.\ O.\ Box 3000, SI-1001 Ljubljana, Slovenia}

\abstract{
The Zee-Babu model is an economical  framework for neutrino mass generation as two-loop quantum corrections. In this work, we present a UV completion of this model by embedding it into an $SU(5)$ unified framework. Interestingly, we find that loop-induced contributions to neutrino masses arising from colored scalars are just as important as those from color-neutral ones. These new states, which are required from gauge coupling unification and neutrino oscillation data to have masses below $\mathcal{O}(10^3)$ TeV, may be accessible to future collider experiments. Additionally, the model can be probed in proton decay searches. Our Markov chain Monte Carlo analysis of model parameters shows a high likelihood of observable $p \rightarrow e^+ \pi^0$ decay signal in the first decade of Hyper-Kamiokande operation. The model predicts a vector-like down-type quark at the TeV scale, utilized for realistic fermion mass generation and gauge coupling unification.
The model is UV-complete in the sense that it is a  unified theory which is realistic and asymptotically free that can be extrapolated to the Planck scale. 
}

\makeatletter
\gdef\@fpheader{}
\makeatother
\begin{document}
\maketitle
\flushbottom

\section{Introduction}

The origin of neutrino masses needed to explain a variety of neutrino oscillation experiments remains a mystery today and is a subject of intense experimental and theoretical scrutiny.  New physics that goes beyond the Standard Model (SM) is required to explain the oscillation data, but the nature of such new physics is completely unknown.  A deeper understanding of its origin could have a profound impact on the evolution of the Universe and may also reveal new symmetries and conservation laws. The seesaw mechanism, which introduces gauge singlet fermions to the SM can elegantly explain the smallness of neutrino masses~\cite{Minkowski:1977sc,Yanagida:1979as,Glashow:1979nm,Gell-Mann:1979vob,Mohapatra:1979ia}; however, the scale of new physics is typically close to the grand unification scale, which cannot be directly tested.  This comment also holds true when gauge non-singlet scalars ($SU(2)_L$ triplets in type-II seesaw~\cite{Mohapatra:1980yp,Schechter:1980gr,Lazarides:1980nt}) or fermions ($SU(2)_L$ triplets in type-III seesaw~\cite{Foot:1988aq}) are introduced to generate small neutrino masses.

An alternative to the high-scale seesaw is the radiative neutrino mass generation mechanism, wherein neutrino mass is zero at tree-level, but is induced via quantum corrections~\cite{Zee:1980ai,Cheng:1980qt,Wolfenstein:1980sy,Zee:1985id,Babu:1988ki} (for reviews, see~\cite{Cai:2017jrq,Babu:2019mfe}). The smallness of neutrino mass in this context arises from loop and chiral suppression factors, and not from the inverse proportionality to a high scale of new physics.  In fact, many realizations of the radiative mechanism require the new physics scale to be close to a TeV, with the potential to be directly tested in collider experiments as well as in flavor observables. 

The goal of this paper is to present an ultraviolet (UV) completion of a simple radiative neutrino mass model, the Zee-Babu model~\cite{Zee:1985id,Babu:1988ki}, by embedding the model into a grand unified theory (GUT) framework~\cite{Pati:1973rp,Pati:1974yy, Georgi:1974sy, Georgi:1974yf, Georgi:1974my, Fritzsch:1974nn} based on the Georgi-Glashow $SU(5)$ gauge symmetry~\cite{Georgi:1974sy}. Such an embedding would be more satisfying and complete, since the new particles needed for neutrino mass generation would belong to complete multiplets of the GUT symmetry group.  Furthermore, the UV-completion we present here is fully realistic, leading to an asymptotically free theory that can be extrapolated all the way to the Planck scale. This includes compatibility with gauge coupling unification, which is problematic in the Georgi-Glashow $SU(5)$ theory~\cite{Georgi:1974sy}, as well as with realistic fermion mass generation, which is also an issue in the original $SU(5)$ theory.  
The unified model presented here is testable in proton decay search experiments, with our Markov chain Monte Carlo  analysis showing a strong likelihood of observing the decay $p \rightarrow e^+ \pi^0$ in the first decade of operation of the Hyper-Kamiokande experiment~\cite{Hyper-Kamiokande:2018ofw}. The model is also testable at colliders through its predictions of a third-family vector-like quark (VLQ), as well as new charged and colored scalars with masses in the TeV--multi-TeV range.

The Zee-Babu model~\cite{Zee:1985id,Babu:1988ki} extends the SM with two new scalars, $\eta^+(1,1,1)$ and $k^{++}(1,1,2)$, with $SU(3)_C \times SU(2)_L \times U(1)_Y$ quantum numbers as indicated. Their Yukawa couplings, $f_{ij} L_i L_j \eta^+ + g_{ij} e^c_i e^c_j k^{--}$, with $f_{ij} = -f_{ji}$ and $g_{ij} = g_{ji}$ ($i,j$ are family indices here), as well as a cubic scalar coupling $\mu \,(\eta^+ \eta^+ k^{--})$ jointly break lepton number by two units, and would generate neutrino Majorana masses via two-loop diagrams involving the exchange of these scalar fields. An $SU(5)$ embedding of these fields can be realized by the addition of a $10_H$ and a $50_H$ to the scalar sector of minimal $SU(5)$ with antisymmetric $10_H$ and symmetric $50_H$ Yukawa coupling matrices, which is what our framework adopts.  Note that the model has no right-handed neutrino $\nu_R$, which goes well with $SU(5)$ unification, which also has no $\nu_R$. The mass relations $m_b^0/m_\tau^0 = m_s^0/m_\mu^0 = m_d^0/m_e^0 = 1$ among the quarks and leptons predicted by the minimal $SU(5)$ model (the superscripts $^0$ indicate GUT scale values of the masses), while a good order of magnitude estimate, fail in detail when compared to experimental values by more than 100\%. These relations can be corrected by introducing a vector-like pair of fermions in the $5+\overline{5}$ representation~\cite{Babu:2012pb}.  Gauge coupling unification is realized in this framework consistently, see Fig.~\ref{fig:GCU}, by choosing the mass scales of certain particles to be below the GUT scale, as listed in Table~\ref{tab:spectrum}. 
A key outcome of our analysis is that, in addition to the original diagrams mediated by $\eta^+$ and $k^{++}$ scalars, neutrinos gain mass through diagrams mediated by their color-triplet and color-sextet counterparts~\cite{Babu:2001ex,Kohda:2012sr,Saad:2020ihm}, see Fig.~\ref{fig:ZBdiagram}. Once the experimental limits from proton decay are combined with gauge coupling unification constraints and perturbativity of Yukawa couplings, we find that both the color-neutral and the colored Zee-Babu states should have masses below $\mathcal{O}(10^3)$ TeV, which can potentially be discovered in future collider experiments, thereby testing the proposed model.

Neutrino mass generation in the framework of $SU(5)$ unification has been studied in conjunction with gauge coupling unification by various authors.  Ref.~\cite{Dorsner:2005fq} has studied type-II seesaw embedding in $SU(5)$, while Ref.~\cite{Bajc:2006ia} has proposed type-III seesaw embedding.  Models of radiative neutrino masses~\cite{Cai:2017jrq,Babu:2019mfe} have also been embedded in $SU(5)$.  The original Zee model~\cite{Zee:1980ai}, where the neutrino masses arise as one-loop radiative corrections, has been embedded in $SU(5)$ in Ref.~\cite{Wolfenstein:1980sf,Barbieri:1981yw,Perez:2016qbo}.  For the embedding of other loop-induced neutrino mass models in $SU(5)$, see for example,  Ref.~\cite{Dorsner:2017wwn,Kumericki:2017sfc,Saad:2019vjo,Dorsner:2019vgf,Dorsner:2021qwg,Antusch:2023jok,Dorsner:2024jiy,Klein:2019jgb,Hinze:2024vrl} (see also Refs.~\cite{Chang:1980ey,deGouvea:2014lva,Boucenna:2014dia,Hagedorn:2016dze}). To the best of our knowledge  the model presented here is the first attempt to embed the Zee-Babu model within $SU(5)$.

This article is organized in the following way. In Sec.~\ref{sec:model}, we introduce the proposed embedding of the Zee-Babu model in $SU(5)$. In-depth discussions on correcting the wrong mass relations for the charged fermions, computation of the new physics contributions to flavor violating processes, and the generation of neutrino masses are provided in Sec.~\ref{sec:mass}. Perturbativity of the Yukawa couplings is examined in Sec.~\ref{sec:perturbativity}. A detailed gauge coupling unification  study, along with proton lifetime analysis, is carried out in Sec.~\ref{sec:unification}. Finally, we conclude in Sec.~\ref{sec:con}.

\section{Embedding the Zee-Babu model in \boldmath{$SU(5)$}}\label{sec:model}
The model we propose extends the  Georgi–Glashow $SU(5)$ model with a $10_H+50_H$ scalar fields and a pair of $5_F+\overline 5_F$ vector-like fermions. The scalars introduced are essential for generating neutrino masses through the Zee-Babu diagrams, while the vector-like fermion pair is crucial in correcting the wrong mass relations among the charged leptons and the down-type quarks~\cite{Babu:2012pb}. Together, these particles will also enable realizing consistent gauge coupling unification.

\textbf{Scalar sector}:--
The decomposition of the scalar fields in the theory under the SM gauge group, $SU(3)_C\times SU(2)_L\times U(1)_Y$, is shown below.
\begin{align}
5_H&=\phi_1(1,2,1/2)+\phi_2(3,1,-1/3),
\\
24_H&=\Phi_1(1,1,0)+\Phi_2(1,3,0)+\Phi_3(8,1,0)+
   +\Phi_4(3,2,-5/6)+\Phi_4^*(\overline{3},2,5/6),
\\
10_H&= \eta_1 (1,1,1)+\eta_2 (\overline{3},1,-2/3)+ \eta_3 (3,2,\frac{1}{6}),  \label{eq:2.3}
\\
50_H&= \chi_1 (1,1,-2) + \chi_2 (3,1,-\frac{1}{3}) + \chi_3 (\overline{3},2,-\frac{7}{6})
+ \chi_4 (6,1,\frac{4}{3}) + \chi_5 (\overline{6},3,-\frac{1}{3}) + \chi_6 (8,2,\frac{1}{2}).\label{eq:2.4)}
\end{align} 
Note that here we denote the original Zee-Babu scalar fields $\eta^+$ and $k^{--}$ as $\eta_1$ and $\chi_1$ respectively, for uniformity of notation.
Among these scalars, only neutral components of the Higgs fields in the adjoint and the fundamental representations acquire vacuum expectation values (VEVs) that break the GUT and the electroweak (EW) symmetries, respectively:
\begin{align}
SU(5) &\xrightarrow[]{\langle 24_H\rangle} SU(3)_C \times SU(2)_L \times U(1)_Y
\\&
\xrightarrow[]{\langle 5_H\rangle} SU(3)_C \times \times U(1)_\mathrm{em} .
\end{align}
The VEVs of these fields are denoted as:
\begin{align}
&
\langle 24_H \rangle =v_{24}\;\mathrm{diag}.(2,2,2,-3,-3), \label{eq:24vev}
\\& \langle 5_H \rangle = (0 \quad 0 \quad 0 \quad 0 \quad v_{5})^T \label{eq:5vev},
\end{align}
with $v_5\simeq 174$ GeV. The nearly degenerate masses of the superheavy gauge boson, $X^{\pm 4/3}_\mu$ and  $Y^{\pm 1/3}_\mu$ are given by  $M_{X,Y}=5\sqrt{2}g_\mathrm{GUT}v_{24}$ (see, for example, Ref.~\cite{Hisano:1992jj}), where $g_\mathrm{GUT}$ represents the unified $SU(5)$ gauge coupling. 

Once the GUT symmetry is broken, the submultiplets of the scalar fields will receive non-degenerate masses. 
In the context of a non-supersymmetric $SU(5)$ GUT, achieving gauge coupling unification while satisfying current proton decay bounds requires a split mass spectrum of the scalar fields.  In the standard $SU(5)$ model, considering the adjoint Higgs, the color octet mass and the $SU(2)$ triplet mass are independent.  See, for example, Eq.~(8) of Ref.~\cite{Babu:2005gx} (see also Ref.~\cite{Guth:1981uk}). If a hierarchical mass spectrum is desired,  one has to make a fine-tuning to get the weak triplet, for example, lighter than the color octet. Since the full Higgs potential has several parameters, it is possible to realize all the hierarchies that are needed for gauge coupling unification.
\color{black} Although there is no mass relations among the submultiplets of $5_H, 24_H$, and $10_H$, there exist three relations within the components of $50_H$~\cite{Dorsner:2014wva,Saad:2019vjo}, which read   
\begin{align}
&m^2_{\chi_4}=3m^2_{\chi_2}-2m^2_{\chi_3},
\\
&m^2_{\chi_5}=2m^2_{\chi_3}-m^2_{\chi_1},
\\
&m^2_{\chi_6}=\frac{3}{2}m^2_{\chi_2}-\frac{1}{2}m^2_{\chi_1}.
\label{eq:relations}
\end{align}
These mass relations will play crucial role in gauge coupling unification, which we discuss in Sec.~\ref{sec:unification}. As for the doublet-triplet mass splitting, the situation in our case is similar to other non-supersymmetric $SU(5)$ models, which is achieved by fine-tuning.

\textbf{Fermion sector}:-- 
As noted before, apart from three families of $\overline 5_F$ and $10_F$ dimensional fermionic representations, the theory contains one pair of  $5_F+\overline 5_F$ vector-like fermions.  These are parametrized as
\begin{align}
\overline{5}_{F}^a=\begin{pmatrix}
d^c_r\\d^c_g\\d^c_b\\e\\ -\nu
\end{pmatrix}_a, \;\;\; 5_{F}=\begin{pmatrix}
d_r\\d_g\\d_b\\e^c\\ -\nu^c
\end{pmatrix}_4, \;\;\;
10^i_F=\frac{1}{\sqrt{2}} \begin{pmatrix}
0&u^c_b&-u^c_g&u_r&d_r\\
-u^c_g&0&u^c_r&u_g&d_g\\
u^c_g&-u^c_r&0&u_b&d_b\\
-u_r&-u_g&-u_b&0&e^c\\
-d_r&-d_g&-d_b&-e^c&0
\end{pmatrix}_i,
\end{align}
where $i=1-3$ and $a=1-4$ are generation indices. Note that $d_4$ is an $SU(2)_L$ singlet while $d_{1,2,3}$ belong to doublets.  The complete Yukawa interactions in our model take the following form:
\begin{align}
-\mathcal{L}_Y&=\frac{1}{4} 10_F^iY^{ij}_{10}10_F^j5_H+\sqrt{2} \overline 5_F^aY^{aj}_510_F^j5^*_H + \overline 5_F^a \underbrace{\left( \mu_a+\eta_a 24_H \right)}_{\equiv \xi_a}5_F^4
\nonumber\\&
+Y^{ab}_A \overline 5_F^a \overline 5_F^b 10_H+Y^{ij}_S  10_F^i 10_F^j 50_H. \label{eq:YUK}
\end{align}
Here, $Y_{10}$ is a $3\times 3$ symmetric matrix, $Y_5$ is $4\times 3$ arbitrary, $\xi_a$ is $4\times 1$ arbitrary,  $Y_S$ is $3\times 3$ symmetric, and  $Y_A$ is $4\times 4$ antisymmetric matrices in flavor space. By overall rotations in flavor space that commute with $SU(5)$ one can bring $Y_5$ to a diagonal form given by
\begin{align}
Y_5^\mathrm{diag}= \begin{pmatrix}
y_1&0&0\\ 
0&y_2&0\\
0&0&y_3\\
0&0&0
\end{pmatrix}, \label{eq:Y5}
\end{align}
while maintaining the symmetry properties of $Y_S$ and $Y_A$. We also define $\hat Y_5^\mathrm{diag}=\mathrm{diag}(y_1,y_2,y_3)$.

\section{Fermion Masses}\label{sec:mass}
\subsection{Charged fermion mass}
Once the electroweak symmetry is broken, the first three terms in Eq.~\eqref{eq:YUK} determine the charged fermion masses, which read 
\begin{align}
-&\mathcal{L}_Y=L^T M_E E^c+ D^TM_D D^c+ u^T M_U u^c,\\
&M_U=v_5Y_{10},
\;\;\;
M_D=\begin{pmatrix}
v_5 \hat Y_5^\mathrm{diag}&0\\
\xi_D^T&\xi_{D_4}
\end{pmatrix}, 
\;\;\;
M_E=\begin{pmatrix}
v_5 \hat Y_5^\mathrm{diag}&\xi_E\\
0&\xi_{E_4}
\end{pmatrix}. \label{eq:massMATRIX}
\end{align}
In the above matrices, we have defined
\begin{align}
&\xi_{D_a}=\mu_a+2\eta_a v_{24},
\;\;\;
\xi_{E_a}=\mu_a-3\eta_a v_{24}, \label{eq:xi}
\\
&\xi_D=\begin{pmatrix}
\xi_{D_1}&\xi_{D_2}&\xi_{D_3}    
\end{pmatrix}^T,\;\;\; \xi_E=\begin{pmatrix}
\xi_{E_1}&\xi_{E_2}&\xi_{E_3}    
\end{pmatrix}^T, 
\end{align}
and the fields 
\begin{align}
&L=(\ell_1,\ell_2,\ell_3,\ell_4)^T, \;\; E^{c}=(e^c_1,e^c_2,e^c_3,e^c_4)^T,
\\
&D=(d_1,d_2,d_3,d_4)^T, \;\; D^{c}=(d^c_1,d^c_2,d^c_3,d^c_4)^T.
\end{align}
A crucial point to note is the difference between  $\xi_D$ and $\xi_E$, arising from the GUT breaking VEV of $24_H$, which helps correct the wrong mass relations of $SU(5)$.

As will be shown later in Sec.~\ref{sec:unification}, from the gauge coupling unification constraints, the vector-like lepton doublet should have a mass near the GUT scale.
Therefore, we integrate it out, which yields the light $3 \times 3$ charged lepton mass matrix and the heavy vector-like lepton mass as~\cite{Babu:1995uu}
\begin{align}
&M_E^\mathrm{light}=\left( \mathds{1}+ \frac{1}{\xi_{E_4}^2}\xi_E \xi_E^\dagger \right)^{-1/2} v_5 \hat Y_5^\mathrm{diag},    \label{eq:Melight}
\\
&M_E^\mathrm{heavy}=\left( \xi_E^\dagger \xi_E + \xi_{E_4}^2 \right)^{1/2} \sim M_\mathrm{GUT}.
\end{align}
 
The vector-like down-type quark, on the other hand, remains at the TeV scale to maximize the GUT scale -- so that proton lifetime limits are satisfied, as discussed in Sec.~\ref{sec:unification}. Hence, for our numerical analysis, we diagonalize the full $4\times 4$ matrix. However, approximate formulas can provide perfect agreement with full numerical solution: 
\begin{align}
&M_D^\mathrm{light}=v_5 \hat Y_5^\mathrm{diag} \left( \mathds{1}+ \frac{1}{\xi_{D_4}^2}\xi_D \xi_D^\dagger \right)^{-1/2},   \label{eq:Mdlight}  
\\
&M_D^\mathrm{heavy}=\left( \xi_D^\dagger \xi_D  + \xi_{D_4}^2 \right)^{1/2} \sim \mathrm{TeV}.
\end{align}
The reason for this excellent agreement is that the VLQs must have masses at least of order TeV, while the light down-type quarks have masses at most above 3 GeV.  Analytic solutions to Eqs.~\eqref{eq:Melight} and~\eqref{eq:Mdlight} showing the consistency of down-type quark and charged lepton masses are also discussed in Ref.~\cite{Babu:2012pb}.

It is conceivable that in presence of the scalar fields $10_H$ and $50_H$, the wrong mass relations of $SU(5)$ among the charged fermions can be corrected via loop diagrams, without the need for the vector-like fermions in the $5_F+\overline 5_F$representation.  While mass relations involving the first two generation quark and leptons can indeed be corrected via loop diagrams involving the exchange of $\eta_3$ and $\chi_3$ fields (Cf: Eq. (\ref{eq:2.3})-(\ref{eq:2.4)})), owing to the antisymmetry of $Y_5$, the third family fermions receive too small a correction to correct for the wrong mass ratio $m_b^0/m_\tau^0 = 1$.  We conclude that the vector-like fermions are essential in this setup for correcting all mass relations.

\subsection{A fit to the charged fermion masses}\label{sec:numerics}

\begin{table}[t!]
\centering
\footnotesize
\resizebox{0.6\textwidth}{!}{
\begin{tabular}{|c|c|c|c|}
\hline
Masses (in GeV)& \pbox{10cm}{~~~~~Inputs \\ (at $\mu= M_{GUT}$)} & \pbox{25cm}{Fitted values\\ (at $\mu= M_{GUT}$)} &pull \\ [1ex] \hline\hline

$m_{d}/10^{-3}$  & 1.14$\pm$0.11 & 1.14&  0.021 \\ \hline
$m_{s}/10^{-3}$  & 21.58$\pm$1.14 & 21.58& 0.064  \\ \hline
$m_{b}$  & 0.994$\pm$0.009 & 0.994&0.066  \\ \hline\hline

$m_{e}/10^{-3}$   & 0.470692$\pm$0.000470 & 0.469544&  0.146 \\ \hline
$m_{\mu}/10^{-3}$  & 99.3658$\pm$0.0993 & 99.2097& 0.193 \\ \hline
$m_{\tau}$  & 1.68923$\pm$0.00168 & 1.68976& 0.042  \\ \hline\hline

$\theta^{\rm{CKM}}_{12}$ &$0.22739\pm0.0006$  &0.2276& 0.100 \\ \hline
$\theta^{\rm{CKM}}_{23}/10^{-2}$ &$4.858\pm0.06$ &0.0486& 0.058 \\ \hline
$\theta^{\rm{CKM}}_{13}/10^{-3}$ &$4.202\pm0.13$ &0.00421& 0.115 \\ \hline
$\delta^{\rm{CKM}}$ &$1.207\pm0.054$ &1.203&0.067  \\ [0.5ex] \hline 
\end{tabular}
}
\caption{Inputs~\cite{Babu:2016bmy} and outputs for the numerical $\chi^2$ analysis.  This benchmark fit with a total $\chi^2=0.1$ shows excellent agreement with data. As an example, for this fit, we additionally demand that the heavy down-type vectorike fermion mass is $m_B \sim 3$ TeV, which safely evades all present collider bounds.}
\label{tab:fit}
\end{table}

Here, we carry out a numerical study showing the consistency of charged fermion masses. In particular, we simultaneously fit the down-type quark and the charged fermion masses obtained from the mass matrices of Eq.~\eqref{eq:massMATRIX}.   For this fit, we take the input values of the masses and mixing parameters at the GUT scale from Ref.~\cite{Babu:2016bmy}, which we summarize in Table~\ref{tab:fit}.

As will be shown later,  gauge coupling unification requires the vector-like down-type quark (denoted as $B$) to have a mass of order the TeV scale. The current LHC bound on the mass of these states is around $m_B \gtrsim 1.5$ TeV~\cite{CMS:2018zkf,ATLAS:2018mpo}.  Therefore, for our benchmark fit, we take its mass to be  $m_B\simeq 3$ TeV,  such that collider constraints are safely satisfied.  The presence of TeV scale vector-like quarks, in general, can mediate flavor violating processes owing to mixing with the light quarks. For the benchmark fit presented below, we find that all flavor violating processes are well under control. In our numerical code, we have checked that the relevant constraints  are satisfied, following those derived in Ref.~\cite{Babu:1988gy,Aguilar-Saavedra:2002phh,Aguilar-Saavedra:2013qpa,Belfatto:2021jhf,Alves:2023ufm}.

Specifically, in our numerical fit procedure, we minimize a $\chi^2$-function defined as
\begin{align}
\chi^2= \sum_i \mathrm{pull}_i^2,\;\;\; \;\;\mathrm{pull}_i= \frac{T_i-O_i}{\sigma_i}.    
\end{align}
Here, $T_i, O_i,$ and $\sigma_i$ represent the theory prediction, experimentally observed central value, and the experimental error associated with the measurements, respectively, for the $i$-th observable.  In the above equation, the sum is taken over all relevant observables to be fitted.  As mentioned above, we fit the down-type quark and charged lepton masses. As will be discussed in the next sub-section, while checking for the flavor violating constraints, one must also need to determine the   Cabibbo–Kobayashi–Maskawa (CKM) matrix. Therefore, in the  $\chi^2$-function, we also include the quark mixing parameters.   The input values in the CKM matrix are summarized in Table~\ref{tab:fit}. Therefore, the sum in the  $\chi^2$-function runs over three down-type quark masses, the mass of the bottom-like heavy quark $m_B$, three charged lepton masses, three mixing angles and the Dirac phase in the quark sector. As for the number of parameters, we have three real and positive couplings $y_i$, three ratios  $\xi_{E_i}/\xi_{E_4}$ (which we take to be real), and  four $\xi_{D_a}$ (for generality, three of which are taken to be complex). As discussed above, we also need to keep $\xi_{D_4}$ as a parameter to fit the mass of the heavy bottom-like quark. Moreover, the parametrization we use for the CKM-like matrix in the up-type quark sector, Eq.~\eqref{eq:ckm}, has three mixing angles and three phases. Hence, we have 13 magnitudes and 6 phases to fit above-mentioned 11 observables (correspondingly, the number of degrees of freedom is 8).    Note that we minimize the $\chi^2$-function to obtain a benchmark parameter set which satisfies all experimental constraints--we have no intention for obtaining the best fit.  This provided benchmark solution has a total $\chi^2=0.1$ (hence, $\chi^2$ per number of degrees of freedom is about 0.01) and pulls of all observables are summarized in Table~\ref{tab:fit}.

A benchmark scenario that reproduces the down-type quarks and charged leptons masses, and satisfies all experimental constraints is given below:
\begin{align}
&\left(y_1,y_2,y_3\right)= \left(1.05\times 10^{-4}, 6.53\times 10^{-4}, 9.77\times 10^{-3}\right), \label{eq:fit1}
\\
&\left(\xi_{E_1},\xi_{E_2},\xi_{E_3}\right)/\xi_{E_4}=   \left(38.506, 22.009, 5.283\right), 
\\
&\left(\xi_{D_1},\xi_{D_2},\xi_{D_3}, \xi_{D_4}\right)=   \left(183.15e^{i 0.049}, 1747.08e^{-i 3.099}, 2459.96e^{i 0.155}, 20.96\right)\mathrm{GeV}. \label{eq:fit2}
\end{align}
Since the vector-like lepton has a mass close to the GUT scale, for the purpose of fitting charged fermion masses, only the ratios $\xi_{E_i}/\xi_{E_4}$ are relevant. However, since the vector-like down-type quark has a mass close to the TeV scale, values of all  $\xi_{D_a}$ are needed to determine its mass as well as for the computation of flavor violating processes.  
The input values of the masses and the fitted values are listed in Table~\ref{tab:fit}. It is to be noted that the input values at the GUT scale given in Table~\ref{tab:fit} assume SM particle spectrum in between $M_Z$ scale and the $M_\mathrm{GUT}$ scale.  In our model, there are additional states that have masses below the GUT scale, which would impact the running of the fermion mass parameters. However, we expect that the  modifications needed to correct for this running  will not be significant. Moreover, any relevant deviations from the values given in Table~\ref{tab:fit} can be incorporated given the freedom we have in this theory.

\subsection{Flavor violation}
For completeness, in the following, we also present the  couplings that are responsible for mediating flavor violating process in the quark sector. 
First, we make a change of basis, 
\begin{align}
D^TM_D D^c\to \overline D_L \hat M_D D_R,\;\;\; \hat M_D=M^\ast_D,    
\end{align}
and diagonalize this $4\times 4$ matrix as
\begin{align}
\hat M_D^{\textrm{diag}}={V^d}_L^\dagger\hat M_D V^d_R.  
\end{align}
We further define
\begin{align}
V^d_L=\begin{pmatrix}
    X_L\\Y_L
\end{pmatrix},   V^d_R=\begin{pmatrix}
    X_R\\Y_R
\end{pmatrix}, 
\end{align}
such that the $X_{L,R}$ are the upper $3\times 4$ blocks from the corresponding $4\times 4$ matrices $V^d_{L,R}$.
A similar analysis provides the matrices $V^u_{L,R}$ and $V^\ell_{L,R}$ in the up-quark and the charged lepton sectors. Furthermore, we  define the mass eigenstates as
\begin{align}
&D_{L,R}=\begin{pmatrix}
d&s&b&B    
\end{pmatrix}^T_{L,R},    
\end{align}
where $B$ represents the bottom-like vector-like quark. As we shall see shortly, this would correspond to a third generation vector-like quark.

With these definitions, the charged current interactions can be written as~\cite{Alves:2023ufm},
\begin{align}
-\mathcal{L}^\mathrm{CC}_W\supset \frac{g}{\sqrt{2}} \overline u_L V\gamma^\mu D_L W^+_\mu + h.c.,    
\end{align}
where $V={V^u_L}^\dagger X_L$ is a $3\times 4$ matrix. Since there is no up-type VLQ,  $V^u_L$ is a $3\times 3$ unitary matrix that diagonalizes the up-quark mass matrix. Therefore, the CKM matrix can be identified with the upper $3\times 3$ block of $V$, i.e., $V_\mathrm{CKM}=V^\mathrm{upper}_{3\times 3}$. During the fitting procedure, we also demand that the CKM matrix is correctly reproduced in a way that all non-unitarity constraints are satisfied. For this purpose, we parametrize  $V^u_L$ as 
\begin{align}
V^u_L=V^u_\mathrm{CKM-like}(\theta_{ij}^u,\delta^u) \mathrm{diag}(e^{i\alpha^u},e^{i\beta^u},1).  \label{eq:ckm}  
\end{align}

The neutral current interactions, on the other hand, take the form~\cite{Alves:2023ufm},
\begin{align}
-\mathcal{L}^\mathrm{NC}_Z\supset \frac{g}{2 \cos\theta_W} \overline D_L F^d\gamma^\mu D_L Z_\mu, 
\end{align}
where we have defined $F^d=X_L^\dagger X_L$. Deviation of $F^d$ from the unit matrix gives rise to flavor changing neutral current (FCNC) processes, such as meson decay and  meson-antimeson oscillations. The coupling responsible for Higgs-mediated FCNC takes the form~\cite{Aguilar-Saavedra:2013qpa}
\begin{align}
-\mathcal{L}_h\supset \overline D_L Y_h D_R \frac{h}{\sqrt{2}}+ h.c., 
\end{align}
here we have defined $Y_h=X_L^\dagger \hat{Y}_5^{\rm diag} X_R$.

Along with the fit parameters given in Eqs.~\eqref{eq:fit1}-\eqref{eq:fit2}, the CKM parameters are correctly reproduced via
\begin{align}
\left(\theta_{12}^u, \theta_{23}^u, \theta_{13}^u, \delta^u, \alpha^u, \beta^u  \right)=  \left(0.765, 0.0613, 0.0347, 0.166, 1.034, -1.175\right).   \label{eq:fit3}
\end{align}
The fitted values of the CKM mixing parameters are also summarized in Table~\ref{tab:fit}.
Eqs.~\eqref{eq:fit1}-\eqref{eq:fit2}, together with Eq.~\eqref{eq:fit3}, let us compute the relevant matrices as:
\begin{align}
&X_L= \left(
\begin{array}{cccc}
 0.54e^{-i 1.52} & 0.83e^{-i 3.09} & 0.0015e^{i 0.05} & 3.65\times 10^{-7}e^{i 1.62} \\
 0.83e^{i 1.61} & 0.54e^{-i 3.09} & 0.092e^{-i 3.09} & 2.1\times 10^{-5}e^{-i 1.52} \\
 0.078e^{-i 1.41} & 0.049e^{i 0.15} & 0.99e^{-i 2.98} & 4.5\times 10^{-4}e^{i 1.72} \\
\end{array}
\right), \label{eq:XL}
\\
&F^d=\left(
\begin{array}{cccc}
 1. & -1.85\times 10^{-9} & 2.46\times 10^{-8}& -5\times 10^{-5} \\
 1.85\times 10^{-9} & 1. & 1.55\times 10^{-8} & 3\times 10^{-5}\\
 -2.46\times 10^{-8} & 1.55\times 10^{-8}  & 1. & -4\times 10^{-4} \\
 -5\times 10^{-5} & -3\times 10^{-5} & 4\times 10^{-4} & 2.10\times 10^{-7} \\
\end{array}
\right), \label{eq:Fd}
\\
&V=   \left(
\begin{array}{cccc}
 0.97e^{-i 2.55} & 0.225e^{i 2.16} & 0.0485e^{-i 1.11} & 2.3\times 10^{-5} e^{i 0.69} \\
 0.225e^{i 2.77} & 0.974e^{-i 1.91} & 0.0042e^{i 2.27} & 4.6\times 10^{-5}e^{-i 0.31} \\
 0.047e^{-i 1.26} & 0.013e^{-i 3.14} & 0.99e^{-i 2.98} & 4.5\times 10^{-4}e^{i 1.72} \\
\end{array}
\right),\label{eq:V}
\\
&Y_h=  \left(
\begin{array}{cccc}
 6.53\times 10^{-6} & 0 & 0 &
   -9.4\times 10^{-4} \\
 0 & 1.2\times 10^{-4} & 0 & 5.9\times 10^{-4} i \\
 0 & 0 & 5.7\times 10^{-3} & -7.8\times 10^{-3} i \\
 -3.54\times 10^{-10} & 0.\,
   -4.25\times 10^{-9} i &
  2.59\times 10^{-6} i
   & 3.65\times 10^{-6} \\
\end{array}
\right). \label{eq:Yh}
\end{align}
As can be seen from these matrices, all flavor violating processes are highly suppressed. Furthermore, Eq.~\eqref{eq:V} depicts that the unitarity of the $3\times 3$ CKM matrix is preserved to  a high degree.  As an example,  we compute the decay $\mathcal{B}_{d,s}\to \mu\overline\mu$. The
SM predictions for such processes are
$BR(\mathcal{B}_d\rightarrow\mu^+\mu^-)=(1.12\pm0.12)\times10^{-10}$ and $BR(\mathcal{B}_s\rightarrow\mu^+\mu^-)=(3.52\pm0.15)\times10^{-9}$~\cite{Blake:2016olu}. The new physics contribution is parameterized by~\cite{Aguilar-Saavedra:2002phh} 
\begin{align}
    \Delta^{\mathcal{B}_q}_{\mu\mu}=-\frac{\pi s^2_W}{\alpha_\text{em}}F^d_{qb},
\end{align}
such that the modified branching ratio is given by
\begin{align}
&    BR\big(\mathcal{B}_q\rightarrow\mu^+\mu^-\big)=         \tau_{\mathcal{B}_q}\frac{G^2_F}{16\pi}\bigg(\frac{\alpha_\text{em}}{\pi s^2_W}\bigg)^2f^2_{\mathcal{B}_q}m_{\mathcal{B}_q}m^2_\mu            
\sqrt{1-\frac{4m^2_\mu}{m^2_{\mathcal{B}_q}}}|\eta^2_Y|\big|
(V_\mathrm{CKM}^*)_{tq}(V_\mathrm{CKM})_{tb}
Y_0(x_t)+\Delta^{\mathcal{B}_q}_{\mu\mu}\big|^2\,.
\end{align}
In the above equation, $\tau_{\mathcal{B}_q}$ and $f_{\mathcal{B}_q}$ are the relevant lifetimes and decay constants, respectively. The factor $\eta_Y$ accounts for the QCD corrections~\cite{Buras:2012ru} and $Y_0$ is one of the well-known Inami-Lim functions~\cite{Inami:1980fz}.   By demanding that the new contribution be smaller than the SM prediction at $95\%$ C.L, the corresponding constraints yield
\begin{align}
&  \big|F^d_{db}\big|<1\times10^{-4}\,, \;\;\;\;\;    \big|F^d_{sb}\big|<4.2\times10^{-4}\,.  
\end{align}
As can be seen from Eq.~\eqref{eq:Fd}, the corresponding $F^d_{ij}$ are about four orders of magnitude smaller. Since the FCNC couplings scale inversely as $m_B^2$, if $m_B \sim 30$ GeV is assumed these modifications would become significant.  However, such low values of $m_B$ are excluded by direct search limits on vector-like quarks.

Finally, using the flavor violating couplings quoted above, one can straightforwardly compute the decay rates of the vectorlike quark~\cite{Aguilar-Saavedra:2013qpa}, and we find
\begin{align}
\Gamma\left( B\to W^-t \right) : \Gamma\left( B\to Z b \right) : \Gamma\left( B\to h b \right) = 2 : 1 : 1 \;,   
\end{align}
as expected from `Higgs boson equivalence theorem'. These are the dominant decay modes of $B$ showing that it can be identified as the third generation vector-like quark (due to $B\to W^-t$, $B\to Z b$, $B\to h b$ dominant decay modes). The corresponding Higgs couplings are given by~\cite{Aguilar-Saavedra:2013qpa}
\begin{align}
&-\mathcal{L}_h\supset \frac{g m_B}{2m_W} 
\begin{pmatrix}
\overline d& \overline s& \overline b    
\end{pmatrix}
\{ Y^L_h P_L + Y^R_h P_R  \} B
h+ h.c., 
\\
&Y^L_h= \begin{pmatrix}
-5.42\times 10^{-5}& 3.41\times 10^{-5} i& -4.54\times 10^{-4} i   \end{pmatrix}^T, 
\\&
Y^R_h= \begin{pmatrix}
-2.04\times 10^{-11} &-2.44\times 10^{-10} i & 1.49\times 10^{-7} i   \end{pmatrix}^T. 
\end{align}
Although they turn out to be subdominant, the model allows us to calculate other branching ratios, such as $W+{\rm jets}$, $h+{\rm jets}$ and $Z+{\rm jets}$ where jets refer to light quarks. For example, from the fits we find that $Br(B \rightarrow W+{\rm jets}) = 0.65\%$, $Br(B \rightarrow Z+{\rm jets}) = 0.5\%$, and $Br(B \rightarrow h+{\rm jets}) = 0.5\%$. We find it interesting that, in principle, GUT scale relations can be tested in decay properties of a surviving light state.

We have explicitly checked numerically that consistent fermion mass fit can be obtained including satisfying all flavor violating processes as long as $m_B\gtrsim 35$ GeV. Below this mass, we find that it violates the $\mathcal{B}_{d,s}\rightarrow\mu^+\mu^-$ decay constraints.  Therefore, 
from the analysis of this section, we see that collider constraints provide the best limit on the mass of the vector-like quark.

\subsection{Neutrino mass generation}

\begin{figure}[t!]
\centering
\includegraphics[width=1\textwidth]{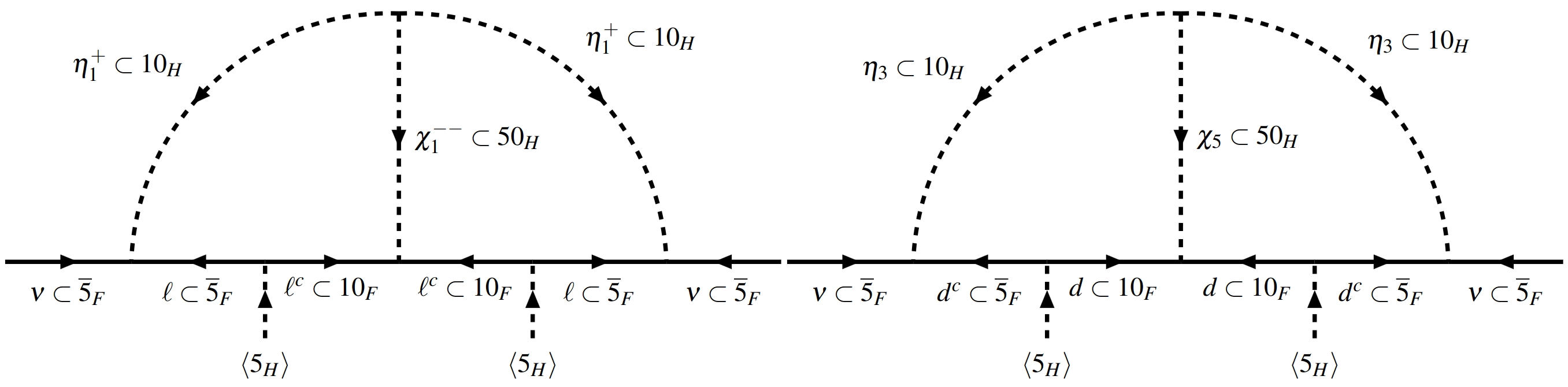}
\caption{Two-loop Feynman diagrams generating Majorana neutrino masses in the Zee-Babu model. Left panel shows the standard Zee-Babu diagram with color-neutral scalars (singly and doubly charged states), while the right panel shows the additional diagram with colored scalars (color-triplet and color-sextet states).  Gauge coupling unification constraints within our unified framework guarantees that both these diagrams must contribute equally to the neutrino mass generation.}
\label{fig:ZBdiagram}
\end{figure}

Neutrino masses are generated via two-loop diagrams in the model as shown  in Fig.~\ref{fig:ZBdiagram}. As can be seen from these figures, there is a Feynman diagram with charged scalars that carry no color ($\eta_1^+$ and $\chi_1^{--}$ of the standard Zee-Babu diagram in the left panel),  as well as a second diagram with colored particles ($\eta_3$ and $\chi_5$, right panel) propagating inside the loop.  As we will show, an interesting consequence of the unified framework is that demanding neutrino mass generation via the Zee-Babu diagram with color-neutral (colored) particles automatically requires that the colored (color-neutral) states must also be light, and therefore, both diagrams contribute equally to the neutrino masses.  

Since in the neutrino mass diagram,  charged leptons and down-type quarks propagate inside loops, we need to compute the neutrino mass formula in the mass diagonal basis of the charged fermions.  Therefore, we diagonalize these matrices,  given in Eq.~\eqref{eq:massMATRIX}, in the following way:
\begin{align}
&\left(M_E\right)_{4\times 4}= U^e_L   M_E^\mathrm{diag}U^{e\dagger}_R,
\;\;\; \left(M_D\right)_{4\times 4}= U^d_L   M_D^\mathrm{diag}U^{d\dagger}_R,
\end{align}
such that
\begin{align}
&L=U^{e*}_L L_0,\;\;\;E^c=U^e_R E^c_0,
\;\;\;
D=U^{d*}_L D_0,\;\;\;D^c=U^d_R D^c_0,
\end{align}
where, $f_0$ represents mass eigenstates. With these rotations, the relevant  interactions for neutrino mass generations can be obtained as  follows:
\begin{align}
\overline 5_F Y_A\overline 5_F 10_H 
&\supset   L^TC Y_A L \eta_1 + L^TCY_Ad^c \eta_3    
\\&
\supset   \nu^TC \underbrace{ \left(U^{e\dagger}_L Y_A U^{e*}_L\right) }_{\equiv \hat Y_A|_{4\times 4}} e_0 \eta_1^+ + \nu^TC \underbrace{\left(U^{e\dagger}_L Y_A U^d_R\right)}_{\equiv \widetilde Y_A|_{4\times 4}} d^c_0 \eta_3^{-1/3},    
\end{align}
Although $\hat Y_A$ remains anti-symmetric, $\widetilde Y_A$ is no longer anti-symmetric. For the latter, this is not required by the SM gauge symmetry.
Furthermore, 
\begin{align}
10_F Y_S 10_F 50_H &\supset   
e^{cT}CY_Se^c\chi_1+Q^TCY_SQ\chi_5
\\&\supset   
e^{cT}_0C  \underbrace{\left(\widetilde U^{eT}_RY_S \widetilde U_R^{e}\right)}_{\equiv \hat Y_S|_{4\times 4}}  e^c_0\chi_1+d^T_0C \underbrace{\left(\widetilde U^{d\dagger}_LY_S \widetilde  U^{d*}_L\right)}_{\equiv \widetilde Y_S|_{4\times 4}} d_0\chi_5.
\end{align} 
Note that both $\hat{Y}_S$ and $\tilde{Y}_S$ remain symmetric, which is required by SM gauge symmetry. Here we have defined $\widetilde U_R^{e}$ to be the upper $3\times 4$ matrix of the $4\times 4$ $U^e_R$ matrix (and same for $U^d_L$). Since $Y_S$ is a $3\times 3$ matrix we needed a change of basis that takes $f^i\to f^a_0$, with $f_a^0$ being the mass eigenstates. All four mass eigenstates run inside the loop of the neutrino diagram (including singly-charged vector-like lepton which has a mass of order the GUT scale).  Thus, $\hat Y_S$ and $\widetilde Y_S$ become $4\times 4$ despite the fact that the original $Y_S$ is a $3\times 3$ matrix. (However, we find that when $E$ runs inside the loop, the diagram is suppressed, see below.)

The $4\times 4$ Majorana neutrino mass matrix from the Zee-Babu diagrams is then given by
\begin{align}
\mathcal{M}^\nu_\mathrm{loop}=\mathcal{M}^{\nu,4\times 4}_{ab}&=16\mu \hat Y_A^{ac} \left(M_E^\mathrm{diag}\right)^{c}\hat Y_S^{cd}\left(M_E^\mathrm{diag}\right)^{d} \hat Y_A^{db}\hat I_{cd}
\nonumber\\&+
48\mu \widetilde Y_A^{ac} \left(M_D^\mathrm{diag}\right)^{c} \widetilde Y_S^{cd}\left(M_D^\mathrm{diag}\right)^{d}  \widetilde Y_A^{db}\widetilde I_{cd}.
\label{4by4} 
\end{align} 
In the above, $I_{cd}$ is the loop function (standing for $\hat{I}_{cd}$ or a $\tilde{I}_{cd}$), given as~\cite{Babu:2002uu,Babu:2015ajp}:
\begin{align}
I_{cd}&= \int \frac{d^4k}{\left(2\pi\right)^4} \int \frac{d^4q}{\left(2\pi\right)^4}   \frac{1}{\left(k^2-m^2_c\right)} \frac{1}{\left(k^2-m^2_{\eta_1}\right)}  \frac{1}{\left(q^2-m^2_d\right)} \frac{1}{\left(q^2-m^2_{\eta_1}\right)} \frac{1}{\left((k-q)^2-m^2_{\chi_1}\right)}.
\end{align}
In $\hat I_{cd}$ ($\widetilde I_{cd}$), $m_i$  represents the masses of the charged leptons (down-type quarks). Moreover, the cubic coupling relevant for neutrino mass generation arises from the following term in the scalar potential:
\begin{align}
V\supset \mu 10_H10_H50_H +h.c. \supset  \mu  \eta_1^-\eta_1^-\chi_1^{--}+ \mu  \eta_3\eta_3\chi_5+h.c. 
\end{align}

Finally, the full $5\times 5$ mass matrix in the neutral fermion sector   is given by
\begin{align}
-\mathcal{L}_Y\supset 
\begin{pmatrix}
\overline{\nu_L}& \overline{\nu^c_R}    
\end{pmatrix} M_N \begin{pmatrix}
\nu_L^c\\ \nu_R 
\end{pmatrix} = \begin{pmatrix}
\overline{\nu_L}& \overline{\nu^c_R}    
\end{pmatrix} 
\begin{pmatrix}
\begin{array}{c|c}
\mathcal{M}^\nu_\mathrm{loop}&\Xi
\\ \hline
\Xi^T&0
\end{array}
\end{pmatrix}_{5\times 5} \begin{pmatrix}
\nu_L^c\\ \nu_R 
\end{pmatrix},    
\end{align}
with
\begin{align}
\Xi^T=\begin{pmatrix}
\xi_{E_1}&
\xi_{E_2}&
\xi_{E_3}&
\xi_{E_4}
\end{pmatrix}.    
\end{align}
In the above, we have defined $\nu_L=\left(\nu_1,\nu_2,\nu_3,\nu_4\right)\subset \overline 5_F^a$. Note that $\nu^c$ and $\bar{\nu^c}=\nu_4$  vector-like neutral leptons are part of $SU(2)_L$ doublets.

We block diagonalize the $M_N$ matrix in the following way:
\begin{align}
M_N=\begin{pmatrix}
\begin{array}{c|c}
\mathcal{M}^{\nu,4\times 4}_\mathrm{loop}&\Xi 
\\ \hline
\Xi^T&0
\end{array}
\end{pmatrix}_{5\times 5} 
 = N_L M^{BD}_N N_L^T,   
\end{align}
such that
\begin{align}
&M^{BD}_N=   \begin{pmatrix}
\begin{array}{c|c}
A_{3\times 3}&0
\\ \hline
0&D_{2\times 2}
\end{array}
\end{pmatrix}_{5\times 5},   
\end{align} 
where we have defined
\begin{align}
&A=   \mathcal{M}^{\nu,3\times 3}_\mathrm{loop},\;\;\;
B=\begin{pmatrix}
\left(\mathcal{M}^{\nu}_\mathrm{loop}\right)_{14}&\xi_{E_1}\\
\left(\mathcal{M}^{\nu}_\mathrm{loop}\right)_{24}&\xi_{E_2}\\
\left(\mathcal{M}^{\nu}_\mathrm{loop}\right)_{34}&\xi_{E_3}
\end{pmatrix},\;\;\;
C=\begin{pmatrix}
\left(\mathcal{M}^{\nu}_\mathrm{loop}\right)_{44}&\xi_{E_4}
\\
\xi_{E_4}&0
\end{pmatrix},
\\
&D=C-B^TA^{-1}B,
\;\;
\mathrm{and,}\;\;
N_L=\begin{pmatrix}
1&0\\
B^TA^{-1}&1
\end{pmatrix}.
\end{align}
Hence, the light neutrino mass matrix, $A=   \mathcal{M}^{\nu,3\times 3}_\mathrm{loop}$, is directly calculated using the formula of Eq.~\eqref{4by4}. In the original Zee-Babu model, the lightest neutrino is massless at two-loop level. In our present model, this is no longer the case since $\mathrm{det}(\mathcal{M}^{\nu,3\times 3}_\mathrm{loop})\neq 0$. However, this result holds true, although approximately.

The neutrino mass contribution from $E$ lepton is highly suppressed, since it scales as $m_\nu \sim (\hat{Y}_A^2 \hat{Y}_S/(16\pi^2)^2) (\mu\,v^2/M_E)$, and since the mass of $E$ is near the GUT scale.  Note that the parameter $\mu$, which also appears in the cubic coupling $\mu\,\eta_1^+ \eta_1^+ \chi_1^{--}$ cannot be much larger than the masses of $\eta_1^+$ and $k^{--}$ owing to perturbative unitarity requirements, has a value in the multi-TeV range. This makes the loop contribution arising from $E$-lepton exchange to be suppressed.  The color-neutral scalar exchange diagram alone would then imply $m_1 \simeq 0$ for the lightest neutrino mass, owing to the determinant of $\hat{Y}_A$ being zero for the upper $3 \times 3$ block.  However, the contributions from the $B$ quark to the neutrino mass are not suppressed, which would imply that $m_1$ is non-zero at the two-loop level.  However, since the $B$-quark contributions are somewhat suppressed owing to the large $B$-quark mass and small mixing with the lighter quarks, we find that $m_1 \ll m_{2,3}$ in the model.

\begin{figure}[t!]
\centering
\includegraphics[width=0.5\textwidth]{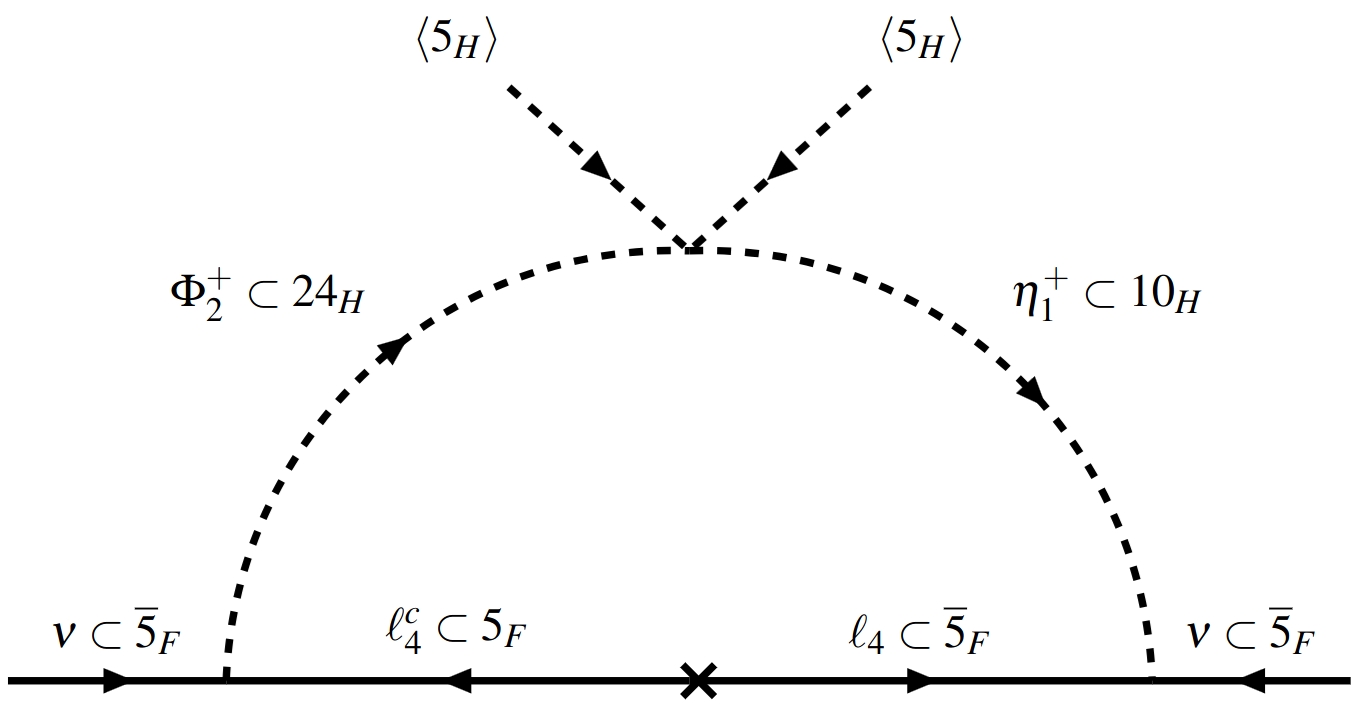}
\caption{Feynman diagram representing Majorana neutrino mass generation at the one-loop order. Within our model, this contribution to the neutrino masses is tiny, see text for details.}
\label{fig:BRPdiagram}
\end{figure}

It is worth mentioning that within this model,  apart from the two-loop Zee-Babu diagram, there is a further contribution to the neutrino masses~\cite{Brdar:2013iea,Kumericki:2017sfc} arising from a one-loop diagram where charged scalars $\Phi^\pm_2$ and $\eta^\pm_1$ run inside the loop along with the vector-like singly charged leptons $\ell^\pm_4$, as shown in Fig.~\ref{fig:BRPdiagram}.
As discussed  in Sec.~{\ref{sec:unification}} unification section, within our scenario, both these states $\Phi_2$ and $L_4$ need to be superheavy (typically close to the GUT scale) to evade experimental bounds on proton decay. Therefore, their contributions to the neutrino mass matrix are found to  be small. The contribution to the  neutrino mass from this one-loop diagram can be estimated as follows~\cite{Brdar:2013iea}:
\begin{align}
\mathrm{M}_\nu&\sim \frac{y_1y_2\lambda}{16\pi^2} \frac{v^2_5\; m_{L_4}}{   m^2_\mathrm{max}},  \;\;\;m_\mathrm{max}=\mathrm{max}\left( m_{L_4}, m_{\Phi_2} \right) 
\\&
\sim (y_1y_2\lambda)  \left(10^{-13}-10^{-14}\right)\mathrm{GeV},
\end{align}
depending on the value of $m_\mathrm{max}$ in the range $(10^{15}-10^{16})$ GeV. Here, $y_{1,2}$ represent the relevant Yukawa couplings and $\lambda$ is the coefficient of the quartic coupling $5^2_H 24_H 10^*_H$ needed to complete the one-loop neutrino mass diagram.

\section{Perturbativity bounds on the Yukawa couplings}\label{sec:perturbativity}
Before moving to the gauge coupling unification analysis, in this section, we study  perturbativity bounds on the Yukawa couplings and find the upper bounds on the Yukawa couplings involving the Zee-Babu scalars. In particular, we choose the maximum values of Yukawa couplings at the low scale demanding that they remain perturbative  (i.e., $y\leq  \sqrt{4\pi}$)  up to the GUT scale.   For this analysis, here we assume that the color-neutral as well as the colored Zee-Babu states all have masses at the TeV scale. Then, the evolution of the gauge and the  Yukawa couplings are governed by the one-loop renormalization group equations (RGEs) given by \begin{align}
16\pi^2 \beta_{g_i}&=16\pi^2 \mu \frac{dg_i}{d\mu}= a_i g^3_i, \;\;\;(a_1,a_2,a_3)=(\frac{83}{15}, \frac{4}{3}, -\frac{25}{6}).    
\end{align}
And the RGEs corresponding to the Yukawa couplings defined as\footnote{Recall that at the GUT scale, Yukawa couplings in the flavor (mass) basis satisfy $Y_h=Y_\eta=Y_A$ ($Y_h=\hat Y_A$, $Y_\eta=\widetilde Y_A$) and $Y_k=Y_\chi=Y_S$ ($Y_k=\hat Y_S$, $Y_\chi=\widetilde Y_S$).} 
\begin{align}
\mathcal{L}\supset  H^* d^c Y_d Q+  H^* e^c Y_e L + u^c Y_u Q H + L Y_h L \eta_1^+ + e^c Y_k e^c \chi_1^{--} +L Y_\eta d^c \eta_3 + Q Y_\chi Q \chi_5,    
\end{align}
take the following form:
\begin{align}
16\pi^2 \beta_{Y_u}&=  Y_u \bigg\{
\frac{3}{2}Y_u^\dagger Y_u -\frac{3}{2}Y_d^\dagger Y_d -\frac{17}{20}g^2_1-\frac{9}{4}g^2_2-8g^2_3 + \mathrm{Tr} \left[ Y_e^\dagger Y_e +3Y_d^\dagger Y_d+3Y_u^\dagger Y_u \right]
\bigg\}
\nonumber\\&
+\bigg\{
6 Y_u Y_\chi^* Y_\chi 
\bigg\},
\\
16\pi^2 \beta_{Y_d}&=  Y_d \bigg\{
\frac{3}{2}Y_d^\dagger Y_d -\frac{3}{2}Y_u^\dagger Y_u -\frac{1}{4}g^2_1-\frac{9}{4}g^2_2-8g^2_3 + \mathrm{Tr} \left[ Y_e^\dagger Y_e +3Y_d^\dagger Y_d+3Y_u^\dagger Y_u \right]
\bigg\}
\nonumber\\&
+\bigg\{
6 Y_d Y_\chi^* Y_\chi + Y_\eta^T Y_\eta^* Y_d
\bigg\},
\\
16\pi^2 \beta_{Y_e}&=  Y_e \bigg\{
\frac{3}{2}Y_e^\dagger Y_e -\frac{9}{4}g^2_1-\frac{9}{4}g^2_2 + \mathrm{Tr} \left[ Y_e^\dagger Y_e +3Y_d^\dagger Y_d+3Y_u^\dagger Y_u \right]
\bigg\}
\nonumber\\&
+\bigg\{
2 Y_eY^\dagger_hY_h + 2 Y_kY_k^*Y_e
+\frac{3}{2} Y_e Y_\eta^* Y_\eta^T
\bigg\},
\\
16\pi^2 \beta_{Y_k}&= \bigg\{
-\frac{18}{5}g^2_1+ 2 \mathrm{Tr}\left[ Y_kY_k^\dagger  \right]
\bigg\} Y_k
+  Y_e Y_e^\dagger Y_k + Y_k Y_e^* Y_e^T  + 4 Y_k Y_k^\dagger Y_k,
\\
16\pi^2 \beta_{Y_h}&= \bigg\{
-\frac{9}{10}g^2_1-\frac{9}{2}g^2_2+ 4 \mathrm{Tr}\left[ Y_hY_h^\dagger  \right]
\bigg\} Y_h
+ \frac{1}{2} Y_hY_e^\dagger Y_e + \frac{1}{2} Y_e^T Y_e^*Y_h + 4 Y_h Y_h^\dagger Y_h
\nonumber\\&
+\frac{3}{2} Y_h Y_\eta^* Y_\eta^T +\frac{3}{2}  Y_\eta Y_\eta^\dagger Y_h,
\\
16\pi^2 \beta_{Y_\eta}&=  Y_\eta \bigg\{
-\frac{13}{20}g^2_1-\frac{9}{4}g^2_2-4g^2_3 + \mathrm{Tr} \left[ Y_\eta Y_\eta^\dagger  \right]
\bigg\}
\nonumber\\&
+\bigg\{
2 Y_h Y_h^\dagger Y_\eta+ Y_\eta Y_d^* Y_d^T + \frac{1}{2} Y_e^T Y_e^* Y_\eta +\frac{5}{2} Y_\eta Y_\eta^\dagger Y_\eta
\bigg\},
\\
16\pi^2 \beta_{Y_\chi}&=  Y_\chi \bigg\{
-\frac{1}{10}g^2_1-\frac{9}{2}g^2_2-8g^2_3 + 2\mathrm{Tr} \left[ Y_\chi Y_\chi^\dagger  \right]
\bigg\}
\nonumber\\&
+\bigg\{
\frac{1}{2}Y_\chi Y_d^\dagger Y_d +\frac{1}{2}Y_\chi Y_u^\dagger Y_u
+ \frac{1}{2} Y_d^T Y_d^* Y_\chi+ \frac{1}{2} Y_u^T Y_u^* Y_\chi
+12 Y_\chi Y_\chi^* Y_\chi
\bigg\}.
\end{align}
We have used the program {\tt SARAH}~\cite{Staub:2008uz} to obtain these RGEs.

Taking these RGEs listed above and turning on each entry of $Y_h$ at a time, we compute the maximum value of the corresponding Yukawa coupling at the low scale. For this evolution, we run these parameters from TeV to the GUT scale, which we fix at $2\times 10^{16}$ GeV. For the analysis the gauge couplings and charged fermion masses are taken as inputs at the 1 TeV scale from Ref.~\cite{Antusch:2013jca}. We repeat the same  procedure for $Y_k, Y_\eta,$ and $Y_\chi$. The results we obtain are summarized below:
\begin{align}
&Y_h^{ij}  \leq 0.56; \;\;\;(\mathrm{antisymmetric}), 
\\
&Y_\eta^{ij}  \leq 1.2,\;\mathrm{except},\;  Y_\eta^{i=j=3}  \leq 2.2,
\\
&Y_k^{i=j}, Y_k^{i\neq j}  \leq 
0.71, 0.62; \;\;\;(\mathrm{symmetric}),
\\
&Y_\chi^{ij}  \leq 
\begin{pmatrix}
0.79& 0.74& 0.73\\
&0.79& 0.73\\
&&0.76    
\end{pmatrix}; \;\;\;(\mathrm{symmetric}).
\end{align}

Using these perturbativity bounds  on the Yukawa couplings, one can estimate the upper bounds on the Zee-Babu states. We obtain this approximate bound knowing that  in the Zee-Babu model, neutrino mass, for the case of normal mass ordering, is dominated by the  (2,2)-entry~\cite{Babu:2002uu} of the doubly charged Yukawa coupling, namely, $Y^{(22)}_k$, and therefore
\begin{align}
m_\nu\sim \frac{1}{16\pi^2} \left( 16 Y_h^2 Y^{(22)}_k m^2_\mu + 48 Y_\eta^2 Y^{(22)}_\chi m^2_s \right)/m_\mathrm{max} \sim 0.05 \;\mathrm{eV}\;\; \Rightarrow m_\mathrm{max}=410 \;\mathrm{TeV}.
\end{align}
Here we have assumed  $\hat I, \widetilde I \sim (16\pi^2)^{-1} m_\mathrm{max}^{-2}$ and $\mu\sim m_\mathrm{max}$.

On the contrary, for the inverted mass ordering, the dominated Yukawa coupling is (2,3)-entry ($Y^{(23)}_k$)~\cite{Herrero-Garcia:2014hfa}. Therefore, replacing $Y^{(22)}_{k,\chi}\to Y^{(23)}_{k,\chi}$ in the above formula, we obtain $m_\mathrm{max}=377 \;\mathrm{TeV}$. 
Accordingly, perturbativity condition on the Yukawa couplings demand the Zee-Babu states to reside $m_\mathrm{max} \lesssim \mathcal{O}(100)$ TeV. Perturbative unitarity suggests~\cite{Babu:2002uu,Herrero-Garcia:2014hfa} that $\mu$ cannot be more than about 3 to 5 times the largest mass running in the loop~\cite{Goodsell:2018tti}. Allowing $\mu\sim 3 m_\mathrm{max}$ gives $m_\mathrm{max}=1230 \;\mathrm{TeV}$ (normal mass ordering), $m_\mathrm{max}=1130 \;\mathrm{TeV}$ (inverted mass ordering).

\begin{figure}[t!]
\centering
\includegraphics[width=0.7\textwidth]{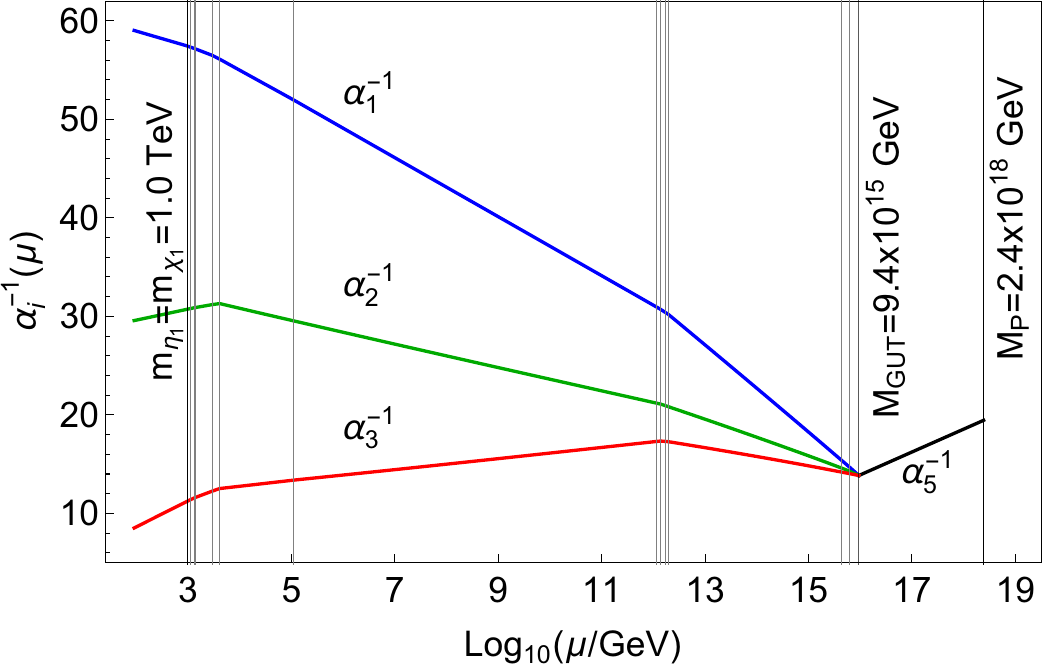}
\caption{An illustration of two-loop gauge coupling unification for a  benchmark scenario given in Table~\ref{tab:spectrum}.}
\label{fig:GCU}
\end{figure}

\section{Gauge coupling unification and proton decay bounds}\label{sec:unification}
In this section, we perform a dedicated study of the gauge coupling unification. 
The two-loop RGEs~\cite{Jones:1981we,Machacek:1983tz,Machacek:1983fi} for the gauge couplings  can be written as
\begin{align}
&\mu\frac{dg_i}{d\mu}=\frac{\beta^{g_i}_{1-\text{loop}}}{16\pi^2} +\frac{\beta^{g_i}_{2-\text{loop}}}{(16\pi^2)^2}\,,
\end{align}
where $\beta^{g_i}_{1-\text{loop}}$ is the one-loop and $\beta^{g_i}_{2-\text{loop}}$ is the two-loop contribution given by
\begin{align}
&\beta^{g_i}_{1-\text{loop}}= \bigg\{ a_i^{\text{SM}}+\sum_X \mathcal{H}(\mu,m_X) \Delta a_i^X \bigg\}  g_i^3\, ,
\\
&\beta^{g_i}_{2-\text{loop}}=  \sum_k  b^{\text{SM}}_{ik}g^2_k +\sum_X \sum_k  \Delta b_{ik}^X g^2_k  \;\mathcal{H}(\mu,m_X),
\end{align}
where the contributions from the Yukawa couplings are ignored and $i,k= 1-3$. The sum is taken over all beyond the SM (BSM) states, $X$.  The Heaviside-Theta function is defined in the following way:
\begin{align}
\mathcal{H}(\mu,m)=
\begin{cases}
1, \;\;\textrm{for}\;\mu\geq m,\\ 
0, \;\;\textrm{for}\;\mu < m.    
\end{cases}
\end{align}
Moreover, $a_i^{\text{SM}}$ and $b^{SM}_{ik}$ are the well known SM one-loop and two-loop $\beta$-coefficients:
\begin{align}
&a_i^{\text{SM}}=(\frac{41}{10}, -\frac{19}{6}, -7),
\\
&b^{\text{SM}}_{ik}=
\left(
\begin{array}{ccc}
 \frac{199}{50} & \frac{27}{10} & \frac{44}{5} \\
 \frac{9}{10} & \frac{35}{6} & 12 \\
 \frac{11}{10} & \frac{9}{2} & -26 \\
\end{array}
\right).    
\end{align}
Whereas, the BSM contributions are listed in the following:
\begin{align}
&\Delta a_i^{\phi_2}=(\frac{1}{15}, 0, \frac{1}{6}),\;\;  
\Delta a_i^{\Phi_2}=(0, \frac{1}{3}, 0),\;\;
\Delta a_i^{\Phi_3}=(0, 0, \frac{1}{2}),\\
&\Delta a_i^{\eta_1}=(\frac{1}{5}, 0, 0),  \;\;
\Delta a_i^{\eta_2}=(\frac{4}{15}, 0, \frac{1}{6}),\;\;
\Delta a_i^{\eta_3}=(\frac{1}{30}, \frac{1}{2}, \frac{1}{3}),\\
&\Delta a_i^{\chi_1}=(\frac{4}{5}, 0, 0),  \;\;
\Delta a_i^{\chi_2}=(\frac{1}{15}, 0, \frac{1}{6}),\;\;
\Delta a_i^{\chi_3}=(\frac{49}{30}, \frac{1}{2}, \frac{1}{3}),\\
&\Delta a_i^{\chi_4}=(\frac{32}{15}, 0, \frac{5}{6}), \;\; 
\Delta a_i^{\chi_5}=(\frac{2}{5}, 4, \frac{5}{2}),\;\;
\Delta a_i^{\chi_6}=(\frac{4}{5}, \frac{4}{3}, 2),\\
&\Delta a_i^{L_4}=(\frac{2}{5}, \frac{2}{3}, 0),  \;\;
\Delta a_i^{d_4}=(\frac{4}{15}, 0, \frac{2}{3}),
\end{align}
and
\begin{align}
&\Delta b_{ik}^{\phi_2}=\left(
\begin{array}{ccc}
 \frac{4}{75} & 0 & \frac{16}{15} \\
 0 & 0 & 0 \\
 \frac{2}{15} & 0 & \frac{11}{3} \\
\end{array}
\right),  \;\;
\Delta b_{ik}^{\Phi_2}=\left(
\begin{array}{ccc}
 0 & 0 & 0 \\
 0 & \frac{28}{3} & 0 \\
 0 & 0 & 0 \\
\end{array}
\right),\;\;
\Delta b_{ik}^{\Phi_3}=\left(
\begin{array}{ccc}
 0 & 0 & 0 \\
 0 & 0 & 0 \\
 0 & 0 & 21. \\
\end{array}
\right),\\
&\Delta b_{ik}^{\eta_1}=\left(
\begin{array}{ccc}
 \frac{36}{25} & 0 & 0 \\
 0 & 0 & 0 \\
 0 & 0 & 0 \\
\end{array}
\right), \;\; 
\Delta b_{ik}^{\eta_2}=\left(
\begin{array}{ccc}
 \frac{64}{75} & 0 & \frac{64}{15} \\
 0 & 0 & 0 \\
 \frac{8}{15} & 0 & \frac{11}{3} \\
\end{array}
\right),\;\;
\Delta b_{ik}^{\eta_3}=\left(
\begin{array}{ccc}
 \frac{1}{150} & \frac{3}{10} & \frac{8}{15} \\
 \frac{1}{10} & \frac{13}{2} & 8 \\
 \frac{1}{15} & 3 & \frac{22}{3} \\
\end{array}
\right),\\
&\Delta b_{ik}^{\chi_1}=\left(
\begin{array}{ccc}
 \frac{576}{25} & 0 & 0 \\
 0 & 0 & 0 \\
 0 & 0 & 0 \\
\end{array}
\right), \;\; 
\Delta b_{ik}^{\chi_2}=\left(
\begin{array}{ccc}
 \frac{4}{75} & 0 & \frac{16}{15} \\
 0 & 0 & 0 \\
 \frac{2}{15} & 0 & \frac{11}{3} \\
\end{array}
\right),
\Delta b_{ik}^{\chi_3}=\left(
\begin{array}{ccc}
 \frac{2401}{150} & \frac{147}{10} & \frac{392}{15} \\
 \frac{49}{10} & \frac{13}{2} & 8 \\
 \frac{49}{15} & 3 & \frac{22}{3} \\
\end{array}
\right),\\\;\;
&\Delta b_{ik}^{\chi_4}=\left(
\begin{array}{ccc}
 \frac{2048}{75} & 0 & \frac{256}{3} \\
 0 & 0 & 0 \\
 \frac{32}{3} & 0 & \frac{115}{3} \\
\end{array}
\right), \;\; 
\Delta b_{ik}^{\chi_5}=\left(
\begin{array}{ccc}
 \frac{8}{25} & \frac{48}{5} & 16 \\
 \frac{16}{5} & 112 & 160 \\
 2 & 60 & 115 \\
\end{array}
\right),\;\;
\Delta b_{ik}^{\chi_6}=\left(
\begin{array}{ccc}
 \frac{36}{25} & \frac{36}{5} & \frac{144}{5} \\
 \frac{12}{5} & \frac{52}{3} & 48 \\
 \frac{18}{5} & 18 & 84 \\
\end{array}
\right),\\
&\Delta b_{ik}^{L_4}=\left(
\begin{array}{ccc}
 \frac{9}{50} & \frac{9}{10} & 0 \\
 \frac{3}{10} & \frac{49}{6} & 0 \\
 0 & 0 & 0 \\
\end{array}
\right),  \;\;
\Delta b_{ik}^{d_4}= \left(
\begin{array}{ccc}
 \frac{4}{75} & 0 & \frac{16}{15} \\
 0 & 0 & 0 \\
 \frac{2}{15} & 0 & \frac{38}{3} \\
\end{array}
\right).
\end{align}

\begin{table}[th!]
\centering
\footnotesize
\resizebox{0.42\textwidth}{!}{
\begin{tabular}{|c||c|}
\hline

multiplet, $X$ & $m_X$ (GeV)\\ [1ex] \hline\hline

$\phi_2(3,1,-1/3)\subset 5_H$&$1.71\times 10^{12}$  \\ \hline\hline
$\Phi_2(1,3,0)\subset 24_H$&$4.38\times 10^{15}$  \\ \hline
$\Phi_3(8,1,0)\subset 24_H$&$1.12\times 10^3$  \\ \hline\hline
$\eta_1(1,1,1)\subset 10_H$&$1.0\times 10^3$  \\ \hline
$\eta_2(\overline 3,1,-2/3)\subset 10_H$&$1.09\times 10^5$  \\ \hline
$\eta_3(3,2,1/6)\subset 10_H$&$1.34\times 10^3$  \\ \hline\hline
$\chi_1(1,1,-2)\subset 50_H$&$1.0\times 10^3$  \\ \hline
$\chi_2(3,1,-1/3)\subset 50_H$&$1.13\times 10^{12}$  \\ \hline
$\chi_3(\overline 3,2,-7/6)\subset 50_H$&$2.98\times 10^3$  \\ \hline
$\chi_4 (6,1,\frac{4}{3})\subset 50_H$&$1.95\times 10^{12}$  \\ \hline
$\chi_5 (\overline{6},3,-\frac{1}{3})\subset 50_H$&$4.10\times 10^3$  \\ \hline
$\chi_6 (8,2,\frac{1}{2})\subset 50_H$&$1.38\times 10^{12}$  \\ \hline\hline
$L_4^c(1,2,-1/2)\subset 5_F$&$6.18\times 10^{15}$  \\ \hline
$d_4(\overline 3,1,1/3)\subset 5_F$&$1.41\times 10^3$  \\ \hline
\end{tabular}
}
\caption{ Mass spectrum associated with the benchmark gauge coupling unification presented in  Fig.~\ref{fig:GCU}. Masses of $\chi_{4,5,6}$ are determined by the mass relations given in Eq. (\ref{eq:relations}). }
\label{tab:spectrum}
\end{table}

\begin{figure}[t!]
\centering
\includegraphics[width=0.9\textwidth]{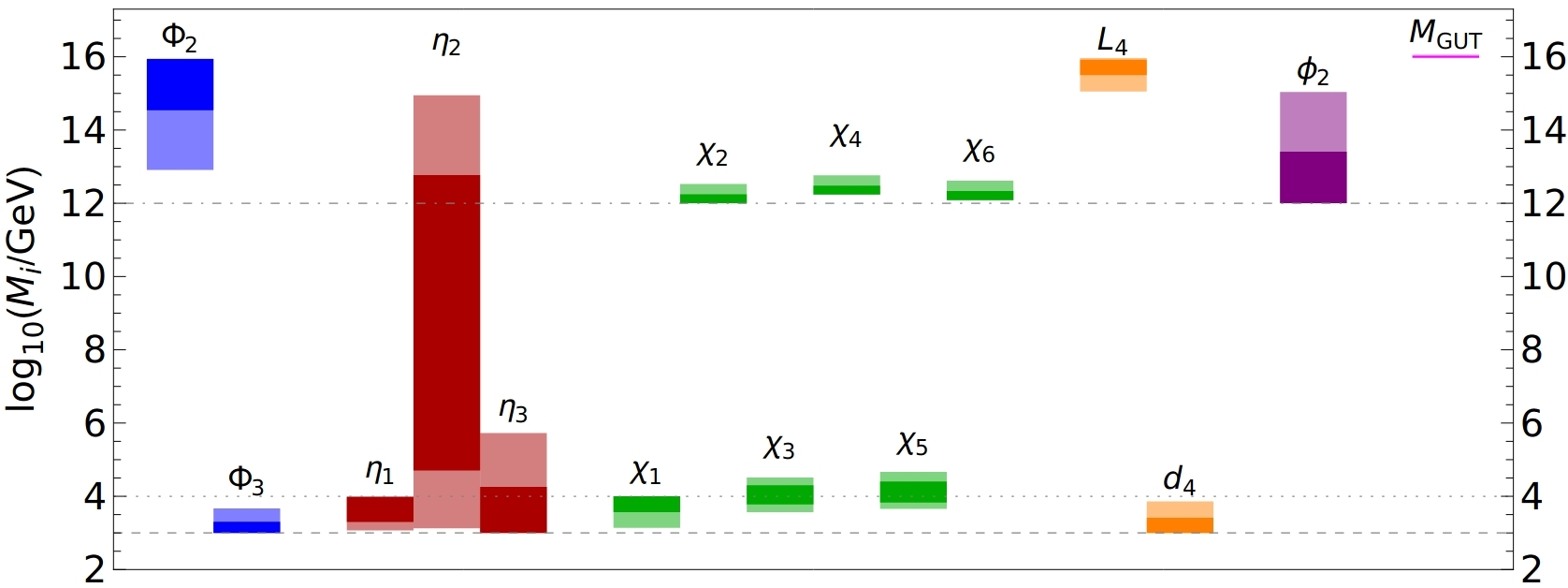}
\caption{Result from extensive Markov chain Monte Carlo (MCMC) analysis.  The $1\sigma$ (dark colored band) and $2\sigma$ (light colored band) highest posterior density (HPD) intervals of the masses of all multiplets from gauge coupling unification constraints. Gray horizontal dashed line represents the approximate collider bound of $m_i\geq 1$ TeV on the states. Gray horizontal dotted line depicts the maximum allowed mass of $m_i\leq 10$ TeV on the $(\eta_1, \chi_1)$ states. Moreover, gray horizontal dashed-dotted line at $\mu= 10^{12}$ GeV scale shows the proton decay bounds on the color-triplet states $\chi_2$ and $\phi_2$.  }
\label{fig:HPD}
\end{figure}

First, a benchmark coupling unification with $m_{h^+}=m_{k^{++}}=1$ TeV is presented in Fig.~\ref{fig:GCU}, with the corresponding particle spectrum summarized in Table~\ref{tab:spectrum}.
Additionally, to explore the full parameter space, 
we perform a detailed Markov chain Monte Carlo (MCMC) analysis for the gauge coupling unification by varying all masses (by imposing the mass relations of Eq. (\ref{eq:relations})) and the result is depicted in Fig.~\ref{fig:HPD}.  For definiteness,  in this analysis, we restrict the Zee-Babu states, namely $\eta_1$ and $\chi_1$ in the range $m_{\eta_1,\chi_1}\in (1,10)$ TeV.  Since the  color-triplet states, $\chi_2$ and $\phi_2$, mediate proton decay, we put a lower bound on their masses of $m_i\geq 10^{12}$ GeV~\cite{Dorsner:2012uz,Dorsner:2024jiy}. In Fig.~\ref{fig:HPD}, the darker (lighter) range represents the $1\sigma$ ($2\sigma$) highest posterior density (HPD) resulting from the MCMC analysis.

Let us briefly discuss the fine-tuning needed, for example, for the case of the  adjoint field. Following Eq.~(8) of Ref.~\cite{Babu:2005gx}, one can write the masses of the weak-triplet, $\Phi_2$, and the  color-octet, $\Phi_3$, as $m^2_{\Phi_2}=\left(-\mu+4\kappa v_{24}\right)v_{24}$ and $m^2_{\Phi_3}=\left(\mu+\kappa v_{24}\right)v_{24}$. Here, $\mu$ and $\kappa$ are relevant (properly re-scaled) coefficients of  cubic and quartic terms in the scalar potential, see Ref.~\cite{Babu:2005gx}. From the above analysis, it is clear that gauge coupling unification demands $m_{\Phi_2}\sim M_\mathrm{GUT}$, while $m_{\Phi_3}\sim$ TeV. Therefore, assuming natural value, in particular, $\mu\sim \mathcal{O}(M_\mathrm{GUT})$ and $\kappa\sim \mathcal{O}(1)$, we accept a fine-tuning of order $\mu+\kappa v_{24}\sim \mathcal{O}(\mathrm{TeV}^2/M_\mathrm{GUT})\sim 10^{-10}$ GeV, leading to $\mu\approx -\kappa v_{24}$.  A similar analysis can be trivially extended for the rest of the multiplets.

Remarkably, the special feature of our model is that both the color-neutral and the colored Zee-Babu states must contribute to neutrino masses. When gauge coupling unification constraints are imposed on top of   demanding neutrino mass generation through  either the color-neutral or the colored Zee-Babu states, the model unambiguously predicts the existence of both colored and color-neutral states around the TeV scale; see Fig.~\ref{fig:HPD}.  Furthermore, consistent gauge coupling unification also calls for the vector-like singlet down-type quark to have a mass of order TeV scale, which may lead to interesting phenomenology. On the other hand, the mass of the vector-like lepton doublet is pushed close to the GUT scale, and therefore, the one-loop Zee-like neutrino mass contribution, as noted before,  is negligible within this part of the parameter space.

The singly ($\eta_1$) and doubly ($\chi_1$) charged scalars, in the standard Zee-Babu model, are expected to have masses close to the TeV scale.  They, therefore, give rise to various interesting experimental signatures that include flavor violating processes as well as unique collider signatures. For experimental probes of these color-neutral states, see for example Refs.~\cite{Babu:2002uu,AristizabalSierra:2006gb,Nebot:2007bc,Ohlsson:2009vk,Schmidt:2014zoa,Herrero-Garcia:2014hfa,Babu:2015ajp,Ruiz:2022sct,Jueid:2023qcf}. As can be seen from Fig.~\ref{fig:HPD}, their colored partners, namely, the color-triplet ($\eta_3$) and color-sextet ($\chi_5$), may also leave enthralling experimental signatures.

\begin{figure}[t!]
\centering
\includegraphics[width=0.7\textwidth]{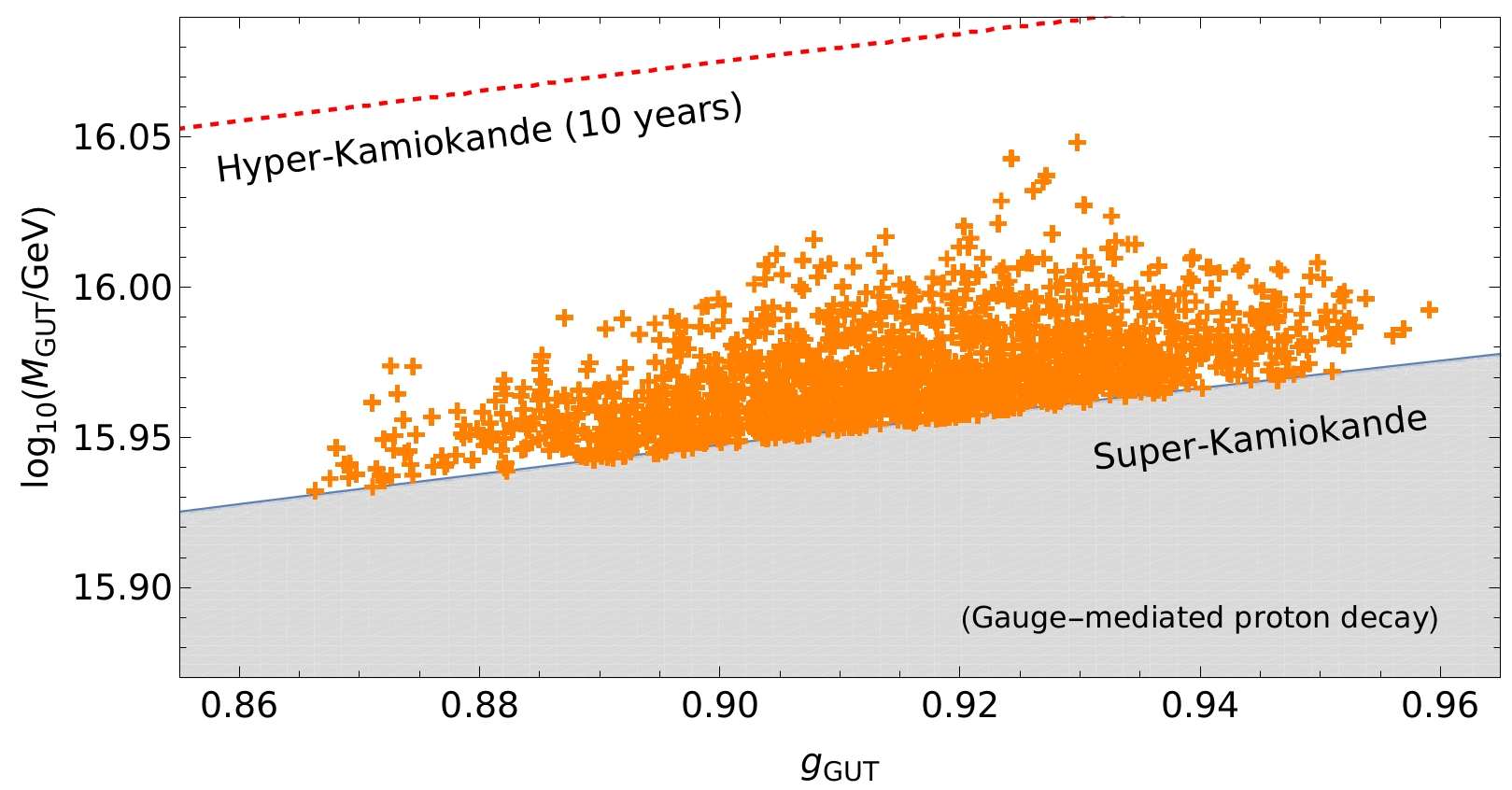}
\caption{Result from the MCMC analysis. Consistent gauge coupling unification points and proton decay bounds in the  $g_\mathrm{GUT}$ vs $M_\mathrm{GUT}$ plane. Current proton decay bound on the decay mode $p\to e^+ \pi^0$ from the Super-Kamiokande is shown by the shaded gray region.  Hyper-Kamiokande sensitivity after its 10 years of operation is depicted with a dashed red line.   }
\label{fig:PD}
\end{figure}

Finally, consistent gauge coupling unification points (from the above mentioned MCMC analysis) and proton decay bounds in the  $g_\mathrm{GUT}$ vs $M_\mathrm{GUT}$ plane is depicted in Fig.~\ref{fig:PD}. The corresponding proton decay bound is computed as follows. The decay rate for the dominant gauge boson mediated decay $p \rightarrow e^+ \pi^0$ are given as~\cite{Nath:2006ut,Babu:2015bna,Dev:2022jbf}
\begin{align}
    \Gamma(p^{+}\to\pi^{0}e^{+})&= \frac{\pi}{4} \frac{m_{p}\,\alpha_{U}^{2}}{f_{\pi}^{2}}\; |\alpha_{H}|^2 R_{\mathrm{L}}^{2} (1+F+D)^2 \left(
    \frac{A_{SR}^{2}}{M_{X}^4}+ \frac{4A_{SL}^{2}}{M_{X}^4}
    \label{eq:PD}
    \right).
\end{align}
Here $f_{\pi}=130.2\,\mathrm{MeV}$ is the pion decay constant $D=0.804$ and $F=0.463$ are chiral Lagrangian parameters ~\cite{Claudson:1981gh,Cabibbo:2003cu}, $\alpha_H=-0.01257\,\mathrm{GeV}^3$ is the hadronic matrix element~\cite{Yoo:2021gql},  $R_L=1.35$ is the long-distance renormalization factor of  the four-fermion operators~\cite{Nihei:1994tx}, $A_{SL}$ and $A_{SR}$ are the short distance factors with $A_{SL} \simeq A_{SR} \simeq 2$~\cite{Babu:2024ecl}, $M_{X}$ is the common mass of the $(X,Y)$ gauge bosons, and $\alpha_{\rm GUT}= g_{\rm GUT}^2/(4\pi)$ is the unified coupling at the GUT scale.  
Note that the value of $\alpha_{\rm GUT} \simeq 1/14$ is significantly larger than the corresponding value in typical non-supersymmetric unified theories where it is $\sim 1/40$. This increased $\alpha_{\rm GUT}$ results in the  shortening of proton lifetime by about an order of magnitude. For the gauge coupling unification study, we have identified the unification scale $M_\mathrm{GUT}$ with the masses of the gauge bosons, i.e., $M_\mathrm{GUT}=M_{X,Y}$. Using Eq.~\eqref{eq:PD}, in Fig.~\ref{fig:PD},  the current (future) proton decay bound of $\tau_p (p\to e^+\pi^0)> 2.4\times 10^{34}$ ($7.8\times 10^{34}$) yrs from Super-Kamiokande~\cite{Super-Kamiokande:2020wjk} (Hyper-Kamiokande~\cite{Hyper-Kamiokande:2018ofw}) is shown with gray shaded region (red dashed line). While, a large part of the parameter space is ruled out by the present experimental limits, intriguingly, 10 yrs operation of Hyper-Kamiokande will likely probe the entire available parameter space of the proposed model.   Although there are more gauge mediated proton decay modes,  the channel $p\to e^+\pi^0$ is the most dangerous and provides the most stringent limits. Additionally,  color-triplet scalars also mediate proton decay. As discussed above, we ensure that all relevant color-triplets have masses $\gtrsim 10^{12}$ GeV (see, for example, Refs.~\cite{Dorsner:2012uz,Dorsner:2024jiy}) so that corresponding experimental limits are satisfied.

\section{Conclusions}\label{sec:con}
The Zee-Babu model stands out as an economical and compelling framework for neutrino mass generation as two-loop radiative correction. In this work, we have proposed a UV-complete version of the Zee-Babu mechanism within the $SU(5)$ framework. A direct consequence of such a unified framework is that in addition to the color-neutral scalars of the Zee-Babu model, there also exist colored GUT partners of these sates. Consistency with gauge coupling unification and neutrino oscillation data requires that both these states contribute equally to the neutrino mass generation mechanism.  We show from perturbativity of the Yukawa couplings and the demand of gauge coupling unification consistent with proton decay bounds that these states should have masses  below $\mathcal{O}(10^3)$ TeV, making them potentially detectable in future collider experiments. The wrong fermion mass relations of the minimal $SU(5)$ theory are corrected in this model by introducing vector-like fermions in the $5+\overline{5}$ representation of $SU(5)$.  Additionally, gauge coupling unification necessitates that the  vector-like down-type quark from this multiplet should have a mass near the TeV scale, which allows for its detection in collider experiments.  Furthermore, our Markov chain Monte Carlo analysis of the model parameters indicates a high likelihood of detecting its signal in proton decay searches during the first decade of Hyper-Kamiokande's operation.

\subsection*{Acknowledgments}
The work of KSB is supported in part by US Department of Energy Grant
Number DE-SC 0016013. 
SS acknowledges the financial support
from the Slovenian Research Agency (research core funding No. P1-0035 and N1-0321). 
The authors acknowledge the Center for Theoretical Underground Physics and Related Areas (CETUP* 2024) and the Institute for Underground Science at Sanford Underground Research Facility (SURF) for hospitality and for providing a conducive environment during the finalization of this work.

\bibliographystyle{style}
\bibliography{reference}
\end{document}